\documentclass{aa} 
\usepackage[utf8]{inputenc}
\usepackage{threeparttable,pifont}
\usepackage{natbib}
\usepackage{caption}

\graphicspath{{images/}}

\begin{document}

\title{\vspace{-.2cm}Closing the gap between Earth-based and interplanetary mission observations: Vesta seen by VLT/SPHERE \thanks{Based on observations made with ESO Telescopes at the Paranal Observatory under programme ID 199.C-0074 (PI: P. Vernazza)}\fnmsep\thanks{Reduced images of Table \ref{tab:ao} are only available in electronic form
at the CDS via anonymous ftp to cdsarc.u-strasbg.fr (130.79.128.5)
or via http://cdsweb.u-strasbg.fr/cgi-bin/qcat?J/A+A/}}

\titlerunning{Vesta seen by VLT/SPHERE}

\author{
          R.~JL.~Fétick\inst{1}
           \and
         L.~Jorda\inst{1}
          \and
         P.~Vernazza\inst{1}
         \and
           M.~Marsset\inst{2,3}
         \and        
          A.~Drouard\inst{1} 
           \and
         T.~Fusco\inst{1,4}
          \and
          B.~Carry\inst{5}     
        \and
           F.~Marchis\inst{6}
          \and
         J.~Hanu{\v s}\inst{7}
         \and
          M.~Viikinkoski\inst{8}
           \and
          M.~Birlan\inst{9}
           \and
         P.~Bartczak\inst{10}
           \and
         J.~Berthier\inst{9}         
         \and          
         J.~Castillo-Rogez\inst{11}              
         \and
         F.~Cipriani\inst{12}                    
         \and
         F.~Colas\inst{9}      
         \and
         G. Dudzi\'{n}ski\inst{10}   
         \and         
         C.~Dumas\inst{13}
         \and
         M.~Ferrais\inst{14}
         \and
          E.~Jehin\inst{14}
         \and           
          M.~Kaasalainen\inst{8}
         \and
         A.~Kryszczynska\inst{10}
            \and
         P.~Lamy\inst{1}
          \and
         H.~Le Coroller\inst{1}  
          \and
         A.~Marciniak\inst{10}
         \and
         T.~Michalowski\inst{10}    
         \and
         P.~Michel\inst{5}
         \and
         L.M.~Mugnier\inst{4}
         \and
         B.~Neichel\inst{1}
        \and
         M.~Pajuelo\inst{9, 15}
        \and
          E.~Podlewska-Gaca\inst{10, 16}
        \and
         T.~Santana-Ros\inst{10}
         \and
         P.~Tanga\inst{5}
         \and
         F.~Vachier\inst{9}     
          \and
          A.~Vigan\inst{1}         
         \and
         O.~Witasse\inst{12}         
         \and         
         B.~Yang\inst{17}
}

   \institute{
           Aix Marseille Univ, CNRS, CNES, Laboratoire d'Astrophysique de Marseille, Marseille, France\\
                 $^*$\email{romain.fetick@lam.fr}
             \and
             Department of Earth, Atmospheric and Planetary Sciences, MIT, 77 Massachusetts Avenue, Cambridge, MA 02139, USA
             \and
            Astrophysics Research Centre, Queen's University Belfast, BT7 1NN, UK 
            \and
            ONERA, The French Aerospace Lab BP72, 29 avenue de la Division Leclerc, 92322 Chatillon Cedex, France
            \and
          Université Côte d'Azur, Observatoire de la Cote d'Azur, CNRS, Laboratoire Lagrange, France
                \and
           SETI Institute, Carl Sagan Center, 189 Bernado Avenue, Mountain View CA 94043, USA 
           \and
         Institute of Astronomy, Charles University, Prague, V Hole\v sovi\v ck\'ach 2, CZ-18000, Prague 8, Czech Republic
            \and
              Department of Mathematics, Tampere University of Technology, PO Box 553, 33101, Tampere, Finland 
         \and
         IMCCE, Observatoire de Paris, 77 avenue Denfert-Rochereau, F-75014 Paris Cedex, France
         \and
         Astronomical Observatory Institute, Faculty of Physics, Adam Mickiewicz University, S{\l}oneczna 36, 60-286 Pozna{\'n}, Poland
      \and
	 Jet Propulsion Laboratory, California Institute of Technology, 4800 Oak Grove Drive, Pasadena, CA 91109, USA
	 \and
	 European Space Agency, ESTEC - Scientific Support Office, Keplerlaan 1, Noordwijk 2200 AG, The Netherlands
	 \and
        TMT Observatory, 100 W. Walnut Street, Suite 300, Pasadena, CA 91124, USA
     \and
         Space sciences, Technologies and Astrophysics Research Institute, Université de Liège, Allée du 6 Août 17, 4000 Liège, Belgium
         	\and
	Sección F\'isica, Departamento de Ciencias, Pontificia Universidad Cat\'olica del Per\'u, Apartado, Lima 1761, Per\'u
         \and
         Institute of Physics, University of Szczecin, Wielkopolska 15, 70-453 Szczecin, Poland
         \and
         European Southern Observatory (ESO), Alonso de Cordova 3107, 1900 Casilla Vitacura, Santiago, Chile
}

\date{July 2018}



\abstract{
Over the past decades, several interplanetary missions have studied small bodies {in situ}, leading to major advances in our understanding of their  geological and geophysical properties. These missions, however, have had a limited  number of targets. Among them, the NASA Dawn mission has characterised in detail the topography and albedo variegation across the surface of asteroid (4)~Vesta down to a spatial resolution of $\sim$20 m/pixel scale.
}{
Here our  aim was to determine how much topographic and albedo information can be retrieved from the ground with VLT/SPHERE in the case of Vesta, having a former space mission (Dawn) providing us with the ground truth that can be used as a benchmark.
}{
We observed Vesta with VLT/SPHERE/ZIMPOL as part of our ESO large programme (ID 199.C-0074) at six different epochs, and deconvolved the collected images with a parametric point spread function (PSF). We then compared our images with synthetic views of Vesta generated from the 3D shape model of the Dawn mission, on which we projected Vesta's albedo information. 
}{  
We show that the deconvolution of the VLT/SPHERE images with a parametric PSF allows the retrieval of the main topographic and albedo features present across the surface of Vesta down to a spatial resolution of $\sim$20--30 km. Contour extraction shows an accuracy of $\sim$1 pixel (3.6~mas). The present study provides the very first quantitative estimate of the accuracy of ground-based adaptive-optics imaging observations of asteroid surfaces.
}{
In the case of Vesta, the upcoming generation of 30-40\,m  telescopes (ELT, TMT, GMT) should in principle be able to resolve all of the main features present across its surface, including the troughs and the north--south crater dichotomy, provided that they operate at the diffraction limit. 
}




\maketitle


\section{Introduction}
\label{sec:intro}

The surface topography of Vesta, the second largest main belt asteroid after the dwarf planet Ceres, has been characterised in detail by the framing camera (FC) \citep{Sierks2011} on board the NASA Dawn mission. The Dawn FC mapped $\sim$80\% of Vesta's surface with image scales of $\sim$20 m/pixel. These images revealed the complex topography \citep{Russell2012, Jaumann2012, Marchi2012, Schenk2012} summarised below: 
\begin{itemize}
    \item The south polar region consists of two overlapping impact basins (Rheasilvia \& Veneneia) and a central mound whose height rivals that of Olympic Mons on Mars. Part of the outer perimeter of the Rheasilvia basin is delimited by a steep scarp. We note that both the impact basin and the central peak were first detected by \citet{Thomas1997} using the Hubble Space Telescope.\\
    \item The   surface is characterised by regions with elevated topography ($\sim$20 km), for example the  Vestalia Terra whose southernmost part merges into the rim area of the Rheasilvia basin \citep{Jaumann2012}.\\
    \item Numerous troughs are present across Vesta's surface, especially in the equatorial and northern regions. Equatorial troughs have lengths that vary from 19 to 380 km and that can be as wide as 15 km, whereas the most prominent northern trough is 390 km long and 38 km wide \citep{Jaumann2012}.\\
    \item Vesta’s cratering record shows a strong north--south dichotomy with Vesta’s northern terrains being significantly more cratered than the southern ones \citep{Marchi2012}.\\
\end{itemize}

The images of the Dawn mission also allowed the production of a high-resolution map of the albedo across Vesta's surface, revealing the greatest variation of normal albedo of any asteroid yet observed; (the normal albedo varies between $\sim$0.15 and $\sim$0.6; \citealt{Reddy2012, Schroeder2014}).\\  

Here we present a new set of ground-based images of Vesta acquired with VLT/SPHERE/ZIMPOL as part of our ESO large programme (ID 199.C-0074; \citealt{Vernazza2018}). These observations,  through direct comparison with the Dawn {in situ} measurements, were performed with the aim of testing  the ultimate resolution achieved for images acquired with this new-generation adaptive-optics (AO) system \citep{beuzit2008sphere,Fusco2006SPIE,Fusco2016SPIE} and  the robustness of our deconvolution algorithm, which uses a synthetic point spread function (PSF) as input. These observations were also used to determine which of the geologic features discovered by Dawn (see above) can already be identified from the ground, and to place a size limit above which these features can be retrieved. Finally, these observations were used to produce an albedo map that could be directly compared with that based on the images of the Dawn FC \citep{Schroeder2014}. In summary, Vesta was used as the benchmark target for our large programme, allowing us to test and ultimately validate our different techniques of image analysis.  


\begin{figure*}[h!]
   \centering
   \includegraphics[width=14cm]{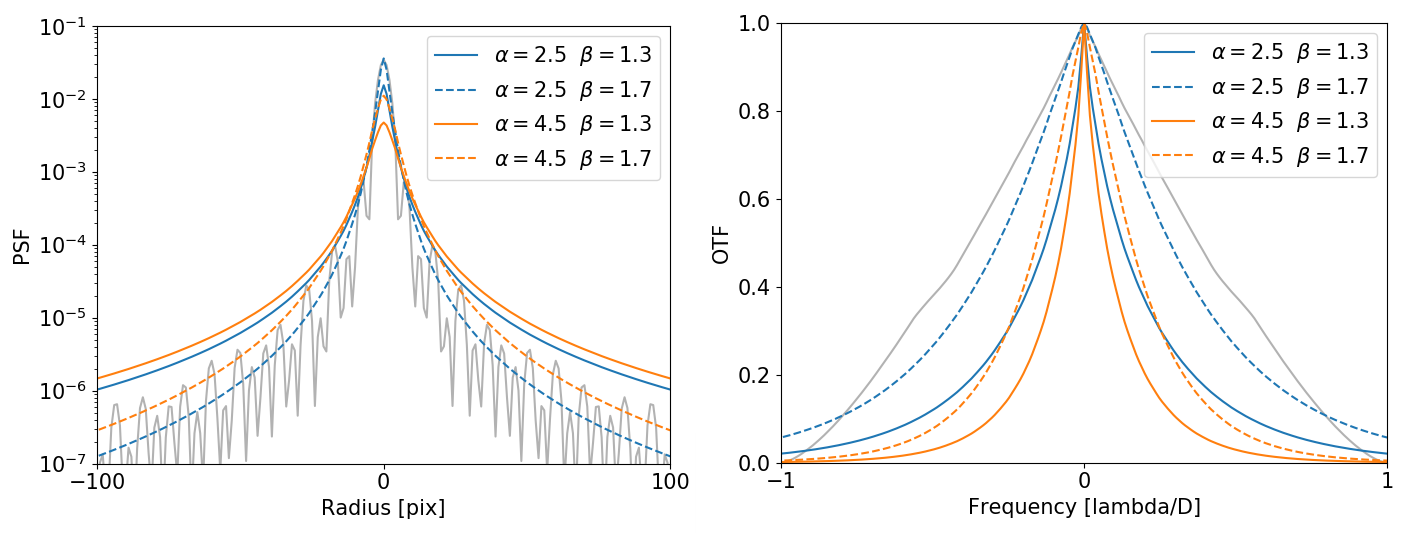}
   \caption{Four of the 25 Moffat PSFs (left) used for deconvolution, and their associated OTF (right). Colours indicate different values of $\alpha$, and line styles show different values of $\beta$. Grey curves in the left and right panels are respectively the PSF and OTF representing the diffraction limit of the SPHERE 8m pupil at $\lambda = 646$ nm.}
\label{fig:PSF_FTO_range}
\end{figure*}

\begin{figure*}[!]
   \centering
   \includegraphics[width=15cm]{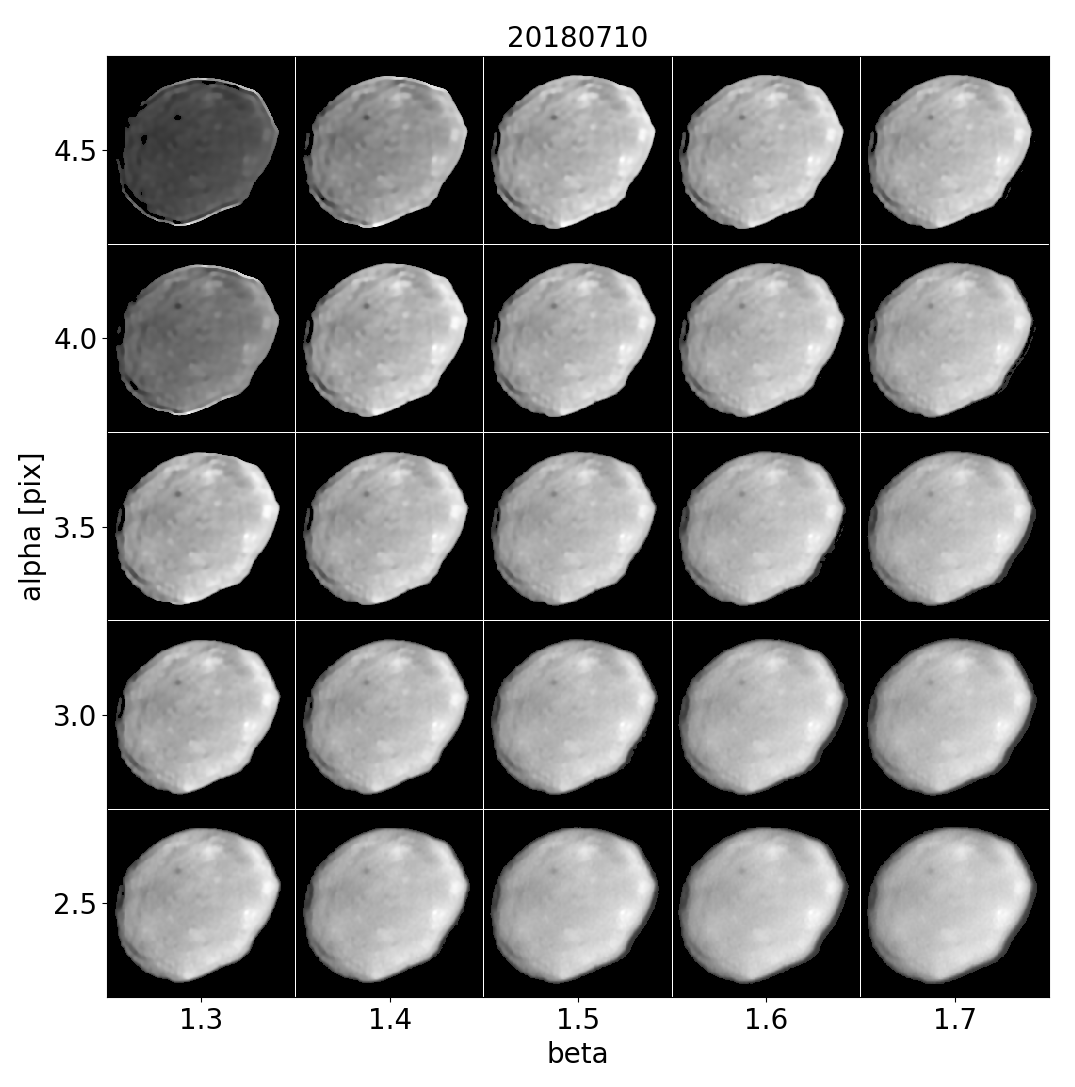}
   \includegraphics[width=14cm]{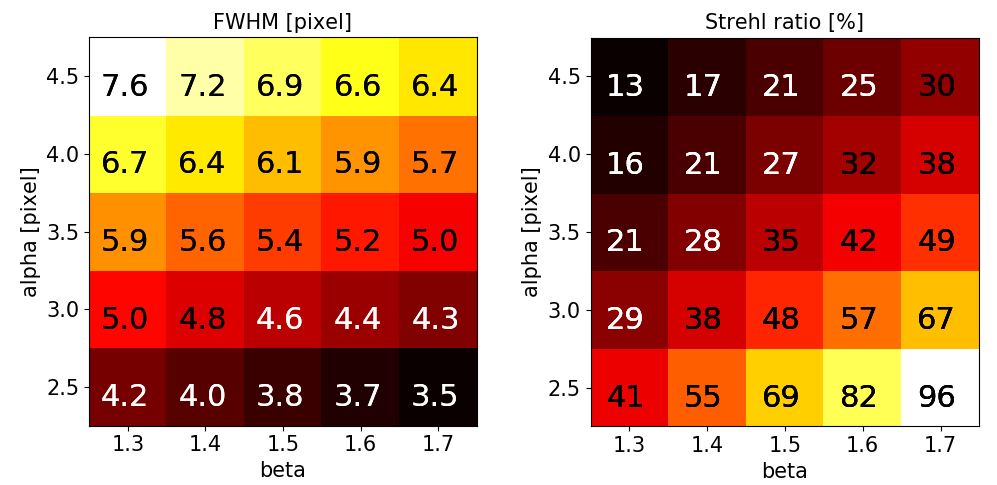}
   \caption{Top: VLT/SPHERE/ZIMPOL observation on 2018 July 10$^\text{th}$ deconvolved using 25 different Moffat parametric PSFs. Each image of Vesta is cropped to $200\times 200$ pixels for best enlargement, and scaled by its maximum intensity for optimum image dynamics. Bottom: Corresponding FWHM and Strehl ratio for the 25 PSFs. Values indicate the FWHM and Strehl ratio respectively, while the colour map (dark for low values, white for high values) helps to visualise the diagonal trend discussed in the text.}
\label{fig:deconv_range}
\end{figure*}

\section{Observations and data reduction}
\label{sec:obs}


We observed (4)~Vesta with the VLT/SPHERE/ZIMPOL instrument \citep{beuzit2008sphere,thalmann2008sphere,schmid2012tests} at six different epochs between May 20, 2018, and July 10, 2018 (see Table \ref{tab:ao} for a complete list of the observations). 
The data reduction protocol is the same for all the targets in our large programme. 
We refer the reader to \citet{Vernazza2018} for a description of the procedure. A subset of the Vesta images after pipeline reduction and before deconvolution are shown in Fig. \ref{fig:VESTA_RAW}. 

\section{Deconvolution method}
\label{sec:deconv}
\subsection{Recurrent deconvolution artefacts with the observed PSF}

Our large programme was initially designed as follows. Each asteroid observation was followed by the acquisition of a stellar PSF, using a star of same apparent magnitude as the asteroid. The asteroid image (cropped to $500\times 500$ pixels) and its associated PSF (same pixel size as the image) were then fed to the MISTRAL algorithm \citep{mugnier2004mistral} for the deconvolution process. It appeared that for nearly $50\%$ of the observations, the deconvolution resulted in strong artefacts at the asteroid edges, highlighted by the presence of a bright corona (see upper left panel of Fig. \ref{fig:deconv_range} for an example). We note that edge issues resulting from the deconvolution process have already been reported in the literature \citep{Marchis2006}.\\

The problem was slightly improved using myopic deconvolution \citep{conan1998myopic} with the MISTRAL algorithm. In this mode, the gradient-descent algorithm estimates simultaneously the couple \{object+PSF\}. Specifically, the PSF observed during the night was used to compute an average optical transfer function (OTF), i.e. the Fourier transform of the PSF, and all the PSFs observed during the large programme were used to compute a variance of the OTF around the average. Using this approach, the artefact intensity was reduced but still visible. This seemed to indicate that the deconvolution issues came from the PSF itself. We further inferred that the deconvolution artefacts may arise from a mismatch between the observed stellar PSF and the true PSF during the asteroid observation. If the shape of the observed PSF is too far from that of the true PSF, the myopic deconvolution cannot completely correct this issue. Moreover, myopic (or blind) deconvolution estimates simultaneously the asteroid image and the PSF, leading to an optimisation process that depends on a large number of parameters. Consequently, the mathematical problem of myopic (or blind) deconvolution might be {degenerated} \citep{Blanco-a-11} and the minimisation algorithm might converge towards the incorrect solution. 
Reducing the number of parameters for the PSF is thus critical to avoid degeneracy and deconvolution issues.\\

\begin{figure*}
   \centering
   \includegraphics[width=18.5cm]{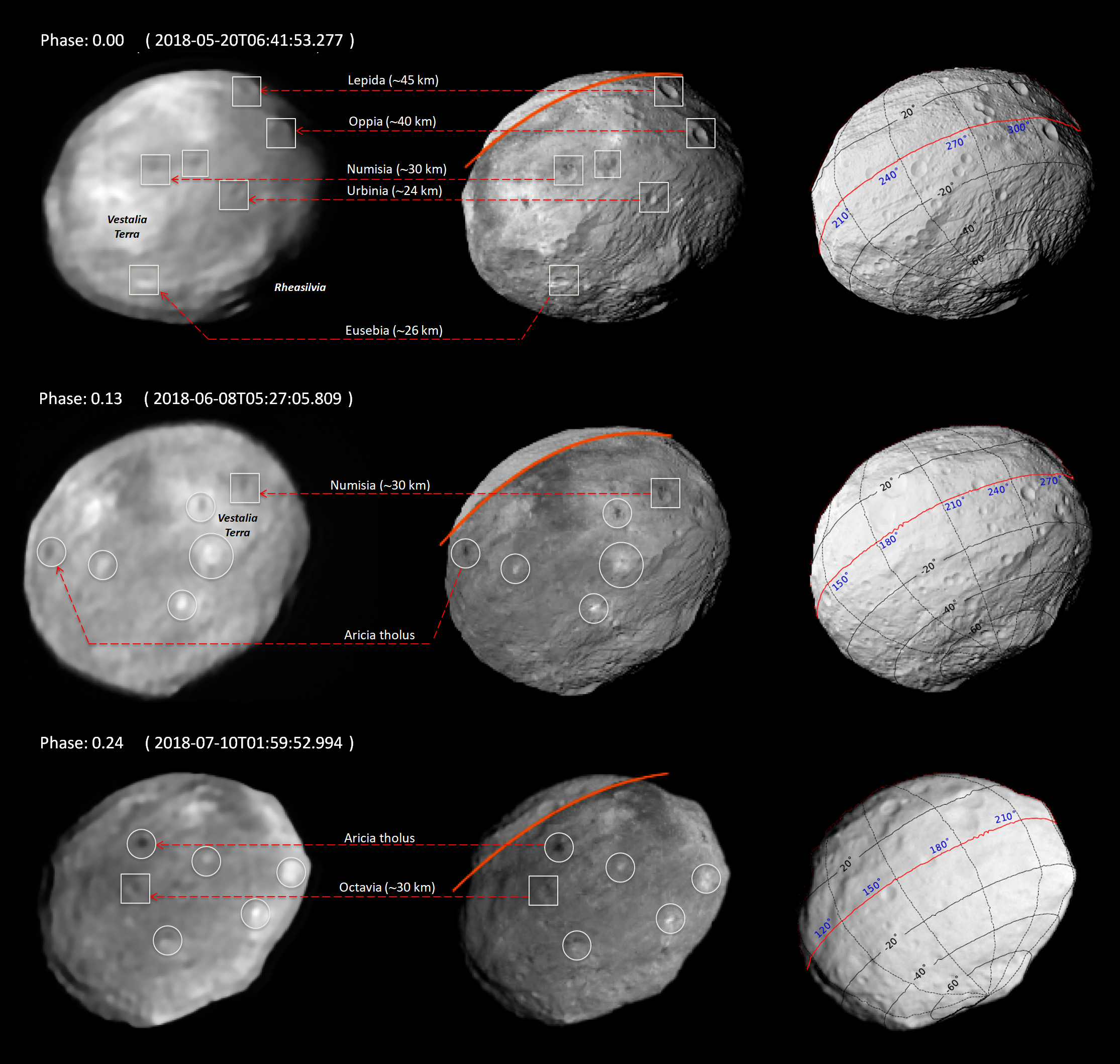}
   \caption{Comparison of the VLT/SPHERE deconvolved images of Vesta (left) with synthetic projections of the Dawn model produced with OASIS and with albedo information from \citet{Schroeder2014} (middle), and the same projection without albedo information but with a latitude/longitude coordinate grid for reference (right). All coordinates are given in the `Claudia' system \citep{Russell2012}. No albedo data is available from Dawn for latitudes above $30^\circ N$ (orange line). Finally, some of the main structures that can be identified in both the VLT/SPHERE images and the synthetic ones are highlighted: craters are embedded in squares and albedo features in circles. The square of the intensity is shown for the left and middle columns to highlight the surface features.}
\label{fig:VESTA_OASIS_A}
\end{figure*}

\begin{figure*}
   \centering
   \includegraphics[width=18.5cm]{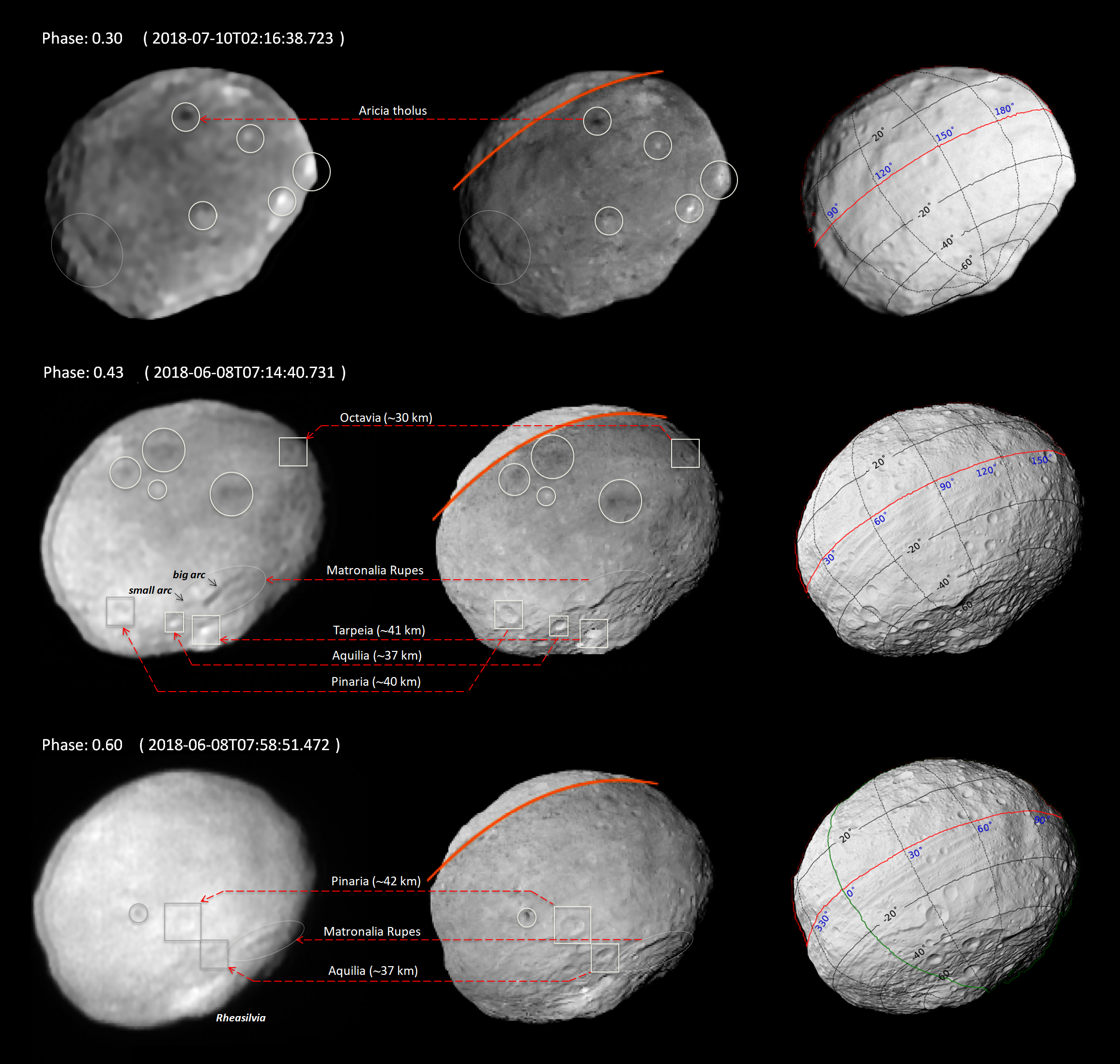}
   \caption*{Fig. 3 \emph{(continued)} Deconvolved images for the two last epochs (phase 0.43 and 0.60) show a clear-dark-clear border on the left. This deconvolution artefact seems to be the outset of the corona artefact discussed in the text.}
\end{figure*}

\subsection{Use of a parametric PSF for the deconvolution}
\label{subsec:parametricPSF}

To overcome the deconvolution issues we faced, we started using parametric PSFs instead of the stellar PSFs. The advantage of a parametric PSF lies in the flexibility of its parameters, which can be adjusted to better estimate the true PSF during our asteroid observation. We chose to model these synthetic PSFs using a Moffat profile \citep{moffat1969theoretical} as such functions are widely used in astronomy to reproduce the sharp coherent peak and wide wings of PSFs in AO observations \citep{andersen2006performance,sanchez2006sinfoni}. Additionally, the Moffat function has  the advantage that it depends on only two parameters. It is defined as 
\begin{equation}
M(r) = \frac{\beta -1}{\pi\alpha^2}\frac{1}{(1+r^2/\alpha^2)^{\,\beta}}
,\end{equation}where $r$ is the radius to the PSF centre, $\alpha$ is a scale factor, and $\beta$ is the Moffat power law index. Setting $\beta>1$ is necessary to have a finite energy. Under this condition, the multiplicative factor $(\beta-1)/\pi\alpha^2$ in the expression ensures that the total PSF energy is equal to unity.\\

As a next step we allowed the two Moffat parameters $\alpha$ and $\beta$ to vary within their realistic range of values:  $(\alpha,\beta)\in~]0,+\infty [~\times~]1,+\infty [$. We then determined a reasonable range of values from a visual inspection of the images: $(\alpha,\beta)\in [2.5,4.5]\times [1.3,1.7]$, where $\alpha$ is given in pixels. Figure \ref{fig:PSF_FTO_range} shows 4 of the 25 Moffat PSFs tested for the deconvolution of our Vesta images (for respectively the minimum and the maximum values of $\alpha$ and $\beta$), and highlights what we consider to be a wide range of PSF shapes for the deconvolution process.\\

After scanning the full range of $\alpha$ and $\beta$, two systematic trends were identified among the deconvolved Vesta images: a large $\alpha$ value in tandem with a small $\beta$ value leads to the corona artefact that regularly occurred using observed PSFs (see 3.1), and  a small $\alpha$ value in tandem with a large $\beta$ value leads to under-deconvolution, meaning that the image blurring is only partially attenuated and the deconvolution is not complete (see Fig. \ref{fig:deconv_range}).

Theses trends can be physically well understood when considering the fact that the full width at half maximum (FWHM) and Strehl ratio of the Moffat PSF are functions of $\alpha$ and $\beta$, where the FWHM can be written as
\begin{equation}
\text{FWHM} = 2\alpha\sqrt{2^{1/\beta}-1}
\end{equation}
and the Strehl ratio as
\begin{equation}
    \text{Strehl ratio} = \frac{M(0)}{A(0)} = \frac{\beta -1}{\pi\alpha^2}\frac{1}{A(0)}
,\end{equation}
with $A(0)$ being the diffraction pattern induced by the pupil (the Airy pattern for a circular non-obstructed aperture) at the centre of the PSF $r=0$.\\ 

The diagonal behaviour observed for the deconvolution with respect to the $\alpha$ and $\beta$ parameters (Fig. \ref{fig:deconv_range}) makes sense when looking at the  FWHM of the corresponding PSF  and Strehl ratios in Fig. \ref{fig:deconv_range} (bottom figure). For large FWHMs (upper left corner)--  or equivalently low Strehl ratios-- deconvolution artefacts appear (corona artefact). This implies that the corona artefacts we encountered during the deconvolution process with observed PSFs most likely occurred either because the observed PSFs were over-estimating the FWHM of our asteroid observations or under-estimating their Strehl ratio. Inversely, for small FWHMs--or large Strehl ratios– the Vesta images appear under-deconvolved.  The diagonal from   bottom left  to  upper right  shows rather stable values for both the FWHM and the Strehl ratio, in agreement with the deconvolution visual quality highlighted in Fig. \ref{fig:deconv_range}. Given the deconvolution trend as a function of the Strehl ratio (Fig. \ref{fig:deconv_range}), we always choose $\alpha$ and $\beta$ parameters such that the object’s edges were as sharp as possible without reaching the point where the bright corona effect appears (Table \ref{tab:param}, Fig. \ref{fig:deconv_range}).  This method has already been applied to the images acquired for (16) Psyche \citep{Viikinkoski2018} and is now systematically applied to all targets within our large programme. In the present case, a quantitative justification (i.e. the accuracy of our deconvolved Vesta images with respect to those of the Dawn mission) for the choice of the parameters $(\alpha,\beta)$  is given in Appendix \ref{sec:appendix_optim}.\\

\begin{table}
\caption{\label{tab:param}Parameters $\alpha$ and $\beta$ chosen for each epoch}
\centering
\begin{tabular}{ccc}
\hline
Epoch & $\alpha$ (pixel) & $\beta$ \\
\hline\hline
2018-05-20 06:41:53 & 4.0 & 1.5 \\
2018-06-08 05:27:05 & 4.0 & 1.5 \\
2018-07-10 01:59:52 & 4.0 & 1.5 \\
2018-07-10 02:16:38 & 4.0 & 1.4 \\
2018-06-08 07:14:40 & 3.0 & 1.5 \\
2018-06-08 07:58:51 & 3.0 & 1.5 \\
\end{tabular}
\end{table}

In summary, the deconvolution with a Moffat PSF converges in practice toward satisfactory results. Within our large programme, we therefore started systematically using a parametric PSF with a Moffat shape to deconvolve our images. The deconvolved images of Vesta are shown in Fig. \ref{fig:VESTA_OASIS_A}.

\begin{figure*}
   \centering
   \includegraphics[width=17cm]{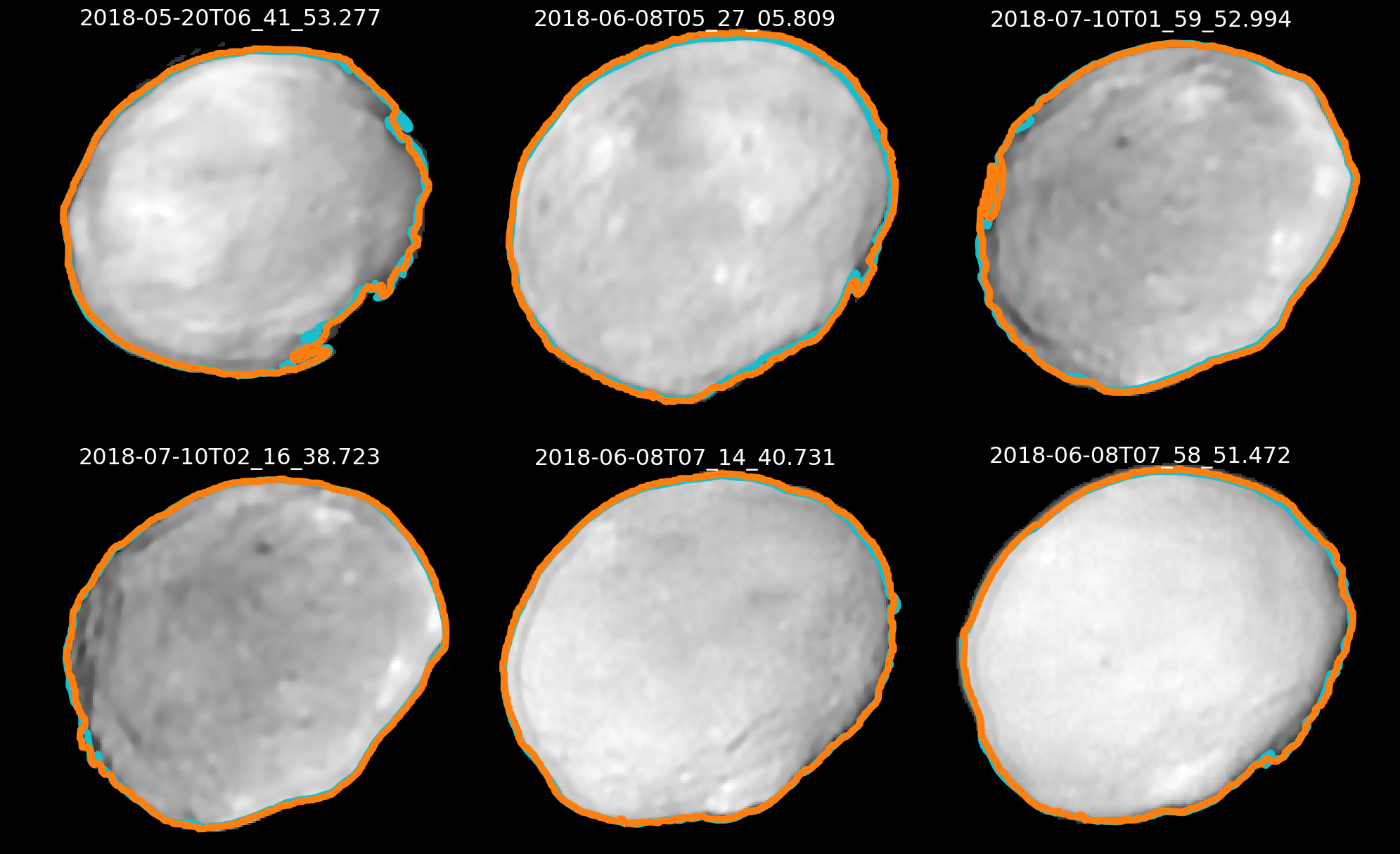}
   \caption{Vesta contours computed for  the synthetic images produced with OASIS (light blue line) and  VLT/SPHERE  after deconvolution (orange line). Contours are plotted over the VLT/SPHERE images for comparison.}
\label{fig:contours}
\end{figure*}

\begin{table*}[t]
    \centering
    \begin{tabular}{c|c|c|c|c|c}
    Epoch (UTC) & $\chi$ [pix] & $\chi_\mathrm{Limb}$ [pix] & $\chi_\mathrm{Term}$ [pix] & $\mathcal{A}_O$ [pix$^2$] & $\mathcal{A}_D$ [pix$^2$] \\
    \hline
    2018-05-20 06:41:53 & 0.83 & 0.41 & 0.99 & 19272 & 19414 (+0.73\%)\\
    2018-06-08 05:27:05 & 1.50 & 0.65 & 1.27 & 22673 & 23382 (+3.12\%)\\
    2018-06-08 07:14:40 & 0.55 & 0.40 & 0.54 & 22762 & 22920 (+0.70\%) \\
    2018-06-08 07:58:51 & 0.60 & 0.42 & 0.71 & 22362 & 22379 (+0.52\%) \\
    2018-07-10 01:59:52 & 1.37 & 0.39 & 1.73 & 21017 & 21224 (+0.98\%)\\
    2018-07-10 02:16:38 & 0.72 & 0.34 & 0.87 & 21223 & 21402 (+0.84\%)\\
    Average $^{\pm\text{std}}$ & $0.93^{\pm 0.37}$ & $0.44^{\pm 0.10}$ & $1.02^{\pm 0.39}$ & & ($+1.15^{\pm 0.89}\%$)\\
\end{tabular}
    \caption{Results on the contour extraction for OASIS and deconvolved images. Deconvolved images considered here are the ones from Fig. \ref{fig:VESTA_OASIS_A} obtained for the $(\alpha,\beta)$ parameters in Table \ref{tab:param}. The $\chi$, $\chi_\mathrm{Limb}$ and $\chi_\mathrm{Term}$ errors are explained in the text. Columns $\mathcal{A}_O$ and $\mathcal{A}_D$ are respectively the area encircled by OASIS and the deconvolution contours. Percentages inside parentheses show the relative error on the estimated deconvolved area with respect to the OASIS area.}
    \label{tab:contour}
\end{table*}

\section{VLT/SPHERE compared to Dawn}
\label{sec:results}


In this section we determine how much topographic and albedo information can be retrieved from the ground with VLT/SPHERE in the case of Vesta, having a former space mission (Dawn) providing us with the ground truth that we can use as benchmark. 

\subsection{OASIS synthetic images}
\label{section:oasis}

The NASA/Dawn in situ mission provided high-resolution images of Vesta used to generate a global shape model \citep{preusker2016}.
The images also provided the rotational parameters of the objects with a high accuracy \citep{konopliv2014}.
Knowing the ephemeris of Vesta and the position of the Earth in the J2000 Equatorial frame, it is thus possible to generate unconvolved synthetic images of Vesta at the time of the SPHERE observations.
We used the  OASIS tool \citep{jorda2010oasis} developed and used in the frame of the Rosetta mission to create these images.
The tool takes into account the pixel scale of the instrument, the viewing and illumination conditions of the object, its global shape, its rotational parameters, and the Hapke parameters describing its bi-directional reflectance properties.
It rigorously accounts for cast shadows and considers the geometric intersection between the triangular facets of the shape model and the pixels to calculate reflectance values for each pixel.
We used the \citet{hapke1986} function built with a one-parameter Henyey-Greenstein phase function and the parameters of \citet{li2013} to describe the reflectance of the surface. 
These synthetic images represent the resolution limit that could have been achieved with a turbulent-less atmosphere and perfect optics. Finally, OASIS images are multiplied by the albedo map from \citet{schroder2014reprint} derived from the Dawn mission and covering the southern part of Vesta up to $30^\circ$N (see Fig. \ref{fig:VESTA_OASIS_A}, middle column).

\begin{figure*}[t]
   \centering
   \includegraphics[width=14cm]{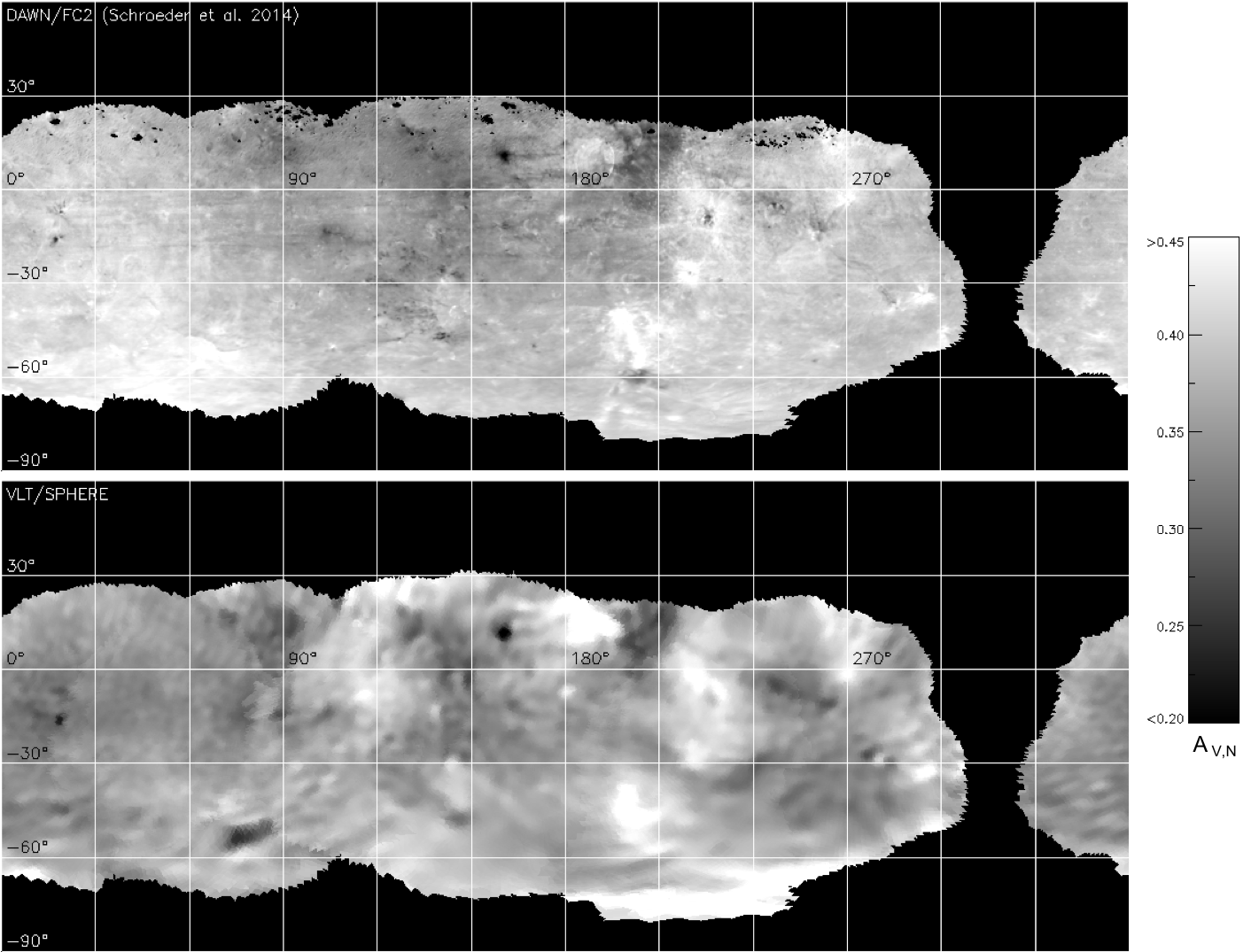}
   \caption{Albedo map of Vesta constructed from VLT/SPHERE images (bottom) compared to the Dawn Framing Camera 2 (FC2) map derived in situ (top; \citealt{Schroeder2014}). The two maps are in equidistant cylindrical projection. 
The Dawn map only shows the region of Vesta covered by SPHERE to facilitate the comparison. The north region of the Dawn map, with the latitude $\phi$ in the $[0^\circ,+30^\circ]$ range, contains several pixels with no available information. Those pixels are left black. }
\label{fig:sph_map}
\end{figure*}

\begin{figure}[t]
   \centering
   \includegraphics[width=9cm]{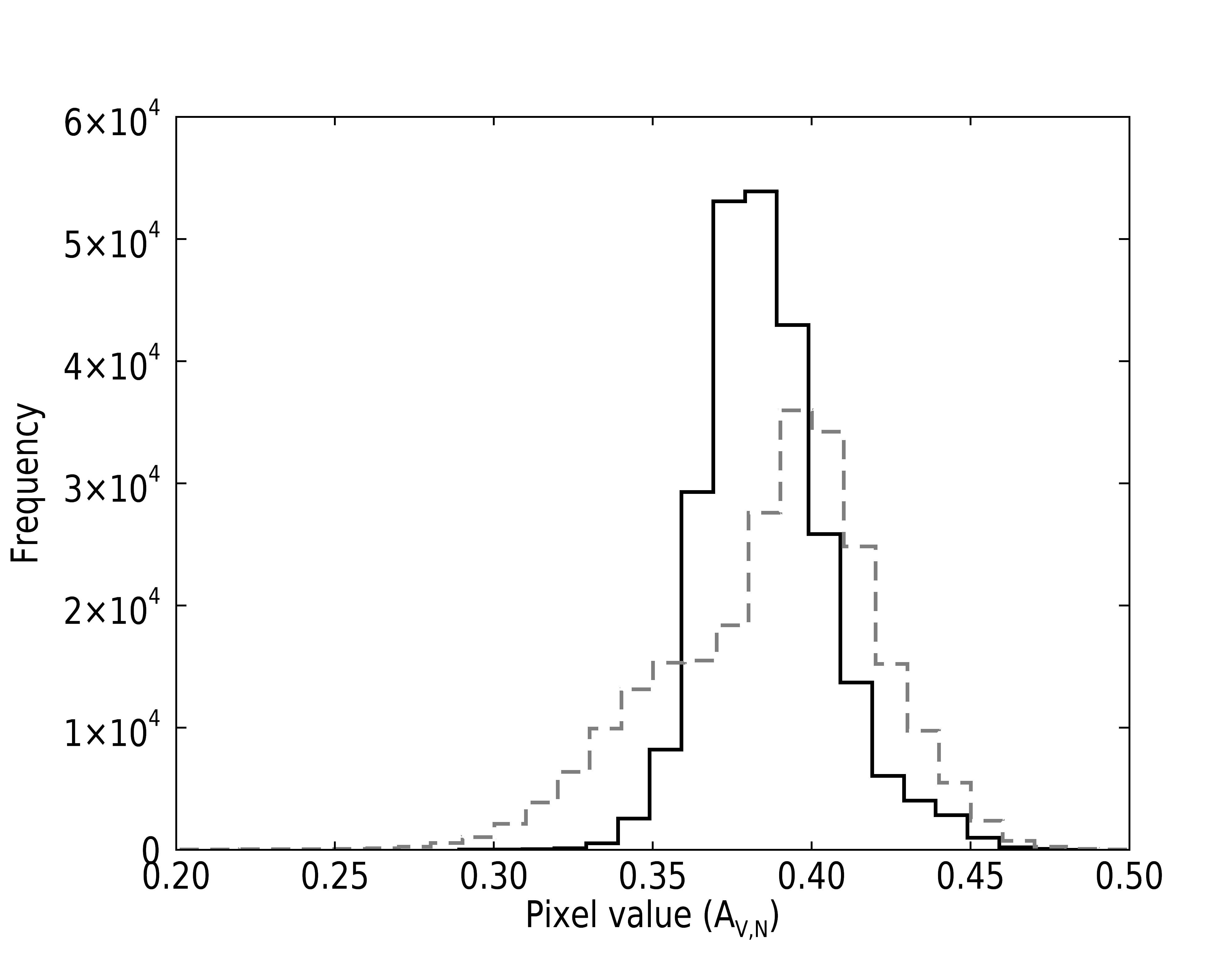}
   \caption{Histogram of pixel values for the SPHERE (continuous black) and Dawn (dotted grey) albedo maps. The SPHERE map exhibits a slightly narrower range of albedo values with respect to Dawn, owing to its lower spatial resolution and residual blurring of the images due to imperfect deconvolution. }
\label{fig:histo_avn}
\end{figure}

\subsection{Contour extraction and comparison}

In order to perform a quantitative comparison between observed and synthetic OASIS images, we   performed a comparison between contour plots. Contours are important for 3D shape reconstruction algorithms. 
The extraction of these contour plots was performed in several steps, described below.
For the observed images, a low threshold $T_1$ was estimated by fitting the histogram of the pixels in a box of $21 \times 21$~pixels around the minimum pixel of the image.
For the synthetic images, the low threshold was set to zero.
A high threshold $T_2$ was then calculated as the maximum pixel value after removing the 10~\% highest pixel values.
The contour level to be used was defined as $T = T_1 + \zeta\ (T_2-T_1)$, where $\zeta$ is a parameter in the interval $[0,1]$ allowing  the value of the contour level to be set with respect to the high and low thresholds. Here $\zeta=0.3$ was chosen.
The image was then converted into a triangular mesh.
For this, the Cartesian coordinates of the vertices were defined by the coordinates $(X=i,Y=j)$ of the pixel $(i,j)$, complemented by its value $Z=P_{ij}$.
Each block of $2 \times 2$ pixels allowed us to define two triangles.
The resulting set of triangles  represents the image as a triangular mesh.
Finally, the intersection of the triangular mesh with the plane $Z=T$ defined the contour plot, represented here as a set of connected 2D points.\\

Let us call $\{D_i\}_{i\in [1,I]}$ and $\{M_j\}_{j\in [1,J]}$ respectively the set of points defining the deconvolved contour and the OASIS model contour. We defined the root mean square metric as
\begin{equation}
\chi = \sqrt{\frac{\sum_i \lambda_i\cdot d(D_i,\{M_j\}_j)^2 }{\sum_i\lambda_i}}
,\end{equation}
where the sum index $i$ runs over all the contour points of the deconvolved image, $d(D_i,\{M_j\}_j)$ is the Euclidean distance between the point number $i$ and its orthogonal projection onto the OASIS contour, $\lambda_i$ is the pixel weighting factor. We chose $\lambda_i$ to be the average distance between $D_i$ and its neighbours $D_{i-1}$ and $D_{i+1}$ as
\begin{equation}
\lambda_i = \frac{1}{2}\left[ d(D_i,D_{i-1}) + d(D_i,D_{i+1}) \right]
\end{equation}
such that an eventual cluster of close points in the contour set does not induce highly localised weighting in the $\chi$ norm. Since the points are nearly homogeneously separated, the correction has little impact on the $\chi$ norm with respect to a simple weighting $\lambda_i=1$. Finally, forcing $\lambda_i = 0$ for some points allows us to select a specific area such as the limb or terminator for contour comparison.\\

Extraction of contours from the synthetic and deconvolved images shows good agreement (see Fig. \ref{fig:contours} for contour visualisation). The match between the contours is better at the limb ($\chi_\text{Limb}=0.44$ pixels) than at the terminator ($\chi_\text{Term}=1.02$ pixels). Indeed the OASIS facets illumination is much more sensitive to angular errors in the orientation of the facets for nearly tangent solar rays  than for normal rays. Moreover, intensity errors in the synthetic images at the terminator translate into larger errors in the contour position because the intensity gradient is smaller there compared to the large jump occurring at the limb.

The average contour error, when taking into account all contour points, is $\chi=0.93$ pixels, showing that sub-pixel contour resolution is achievable on asteroid deconvolved images.\\

The area encircled by the deconvolved contour is larger than the OASIS contour with a relative area difference of $+1.15\%$. This error might be due to residual blurring that was not fully removed by the deconvolution, and to difficulties in extracting the OASIS contour near the terminator. The resulting error on the volume would be $\sim +1.73\%$. All results concerning the contours are summarised in Table \ref{tab:contour}.

\subsection{Reconnaissance of Vesta's main topographic features}

It appears that most of the main topographic features present across Vesta's surface can already be recognised from the ground (Fig. \ref{fig:VESTA_OASIS_A}). This includes the south pole impact basin and its prominent central mound, several D$\geq$25 km sized craters and Matronalia Rupes including its steep scarp and its small and big arcs \citep{Krohn2014}. From these observations, we can determine a size limit of $\sim$30 km for the features that can be resolved with VLT/SPHERE (i.e. features that are 8-10 pixels wide). This detection limit should be in principle sufficient to recognise the north--south crater dichotomy detected by Dawn \citep{Marchi2012, Vincent2014}, according to the fact that the northern hemisphere hosts 70\% of the craters with D$\geq$30 km \citep{Liu2018}. There are several plausible explanations as to why we were not able to resolve this dichotomy. First, we mostly imaged the southern hemisphere, due to Vesta's spin axis orientation. Second, the atmospheric blurring (not entirely removed by the AO) is another limiting factor. Third, the albedo variation across the surface is large, leading to a confusion between the shade of the craters and local albedo variation. This is clearly highlighted in the synthetic images generated by OASIS with and without albedo information (see middle and right column in Fig. \ref{fig:VESTA_OASIS_A}). Fourth, we do not observe the equatorial troughs of Vesta, which are 15 km wide and thus below the detection limit of our images. It follows that future generation telescopes with mirror sizes in the 30-40 m range should in principle be able to resolve the main features present across Vesta's surface, provided that they operate at the diffraction limit.

Finally, we attempt to constrain the detection rate for the largest craters with diameter greater than 40 km.  Except the large Rheasilvia basin, Vesta hosts 21 such craters \citep{Liu2018} and 9 of them were covered by our observations. Of these nine craters, we could identify seven of them (see Fig. \ref{fig:VESTA_OASIS_A}) in our images, implying a detection rate of $\sim$80 \%. The two missing craters are Marcia and Calpurnia, which are part of the snowman crater chain that is located at $\sim13^{\circ}$ north latitude and $\sim195^{\circ}$ eastern longitude. The non-detection of the snowman crater chain is not surprising when looking at the two synthetic OASIS images (with and without albedo information) as the crater chain can barely be recognised in the synthetic images with albedo information (see Fig.~\ref{fig:sph_map}).

\subsection{Reconnaissance of Vesta's main albedo features}

To further test the reliability of the observation and deconvolution procedure, we built an albedo map of Vesta from the deconvolved SPHERE images, and compared this map to that of \citet{Schroeder2014} from {in situ} Dawn measurements. 
We used a subsample of 19 high-quality images from the six epochs of VLT/SPHERE observations.  

Owing to the limited number of geometrical views probed by SPHERE, it is not always possible to differentiate shadows from true albedo variations without prior knowledge about the local surface topography. 
This information is very well constrained in the case of Vesta, and can be retrieved from the OASIS model. 
However, this is not true for most targets in our observing programme, and for almost all asteroids in general. 
We therefore purposely did not use any prior information on Vesta's topography when building the map in order to evaluate our ability to retrieve albedo information from the SPHERE observations alone. 

First, we corrected the illumination gradient present in the SPHERE images, which depends on the local incidence, reflection, and phase angle.
This was performed by fitting a second-order polynomial surface to the disc (image intensity) of Vesta.
This method provides satisfactory results for Earth-based observations of asteroids \citep{Carry2008, Carry2010a}.
Across our observing programme, we also found that it provides better results than using a scattering law (Lommel-Seeliger, Hapke; \citealt{Li2015}) when little is known about the local topography of the object.
Using one of the above-mentioned scattering laws with a low-resolution shape model indeed results in artificially `paved' images owing to the resolution, i.e. the size of the facets, in the model. 
The use of a polynomial fit, which qualitatively mimics the Lommel-Seeliger laws, to the disc  of the object results in smoother photometric correction without artificial discontinuities.
However, it cannot properly fit the illumination of local terrain features, such as craters and mountains.

For each SPHERE image, we then defined a region of interest (ROI) containing the set of pixels to be projected on the map. 
As discussed in Section~\ref{sec:deconv}, an over-deconvolution usually enhances the brightness of the image regions with strong luminosity gradient, such as the asteroid border. 
To avoid including pixels affected by this effect, we only considered  pixels contained in the central region of the asteroid, i.e. 20 pixels away from the asteroid contour (limb and terminator). 

Next, the longitude and latitude of each pixel contained in the ROI was measured using a projection of the OASIS model (Section~\ref{section:oasis}). 
For other targets in our programme, this would be performed using a lower resolution model derived with the ADAM software \citep{Viikinkoski2015}. 

Each pixel value was then projected on the map, using an equidistant cylindrical projection. 
The individual maps from the different epochs of observation were then combined  using the overlapping regions to balance their brightness level. Specifically, the combined map was calculated as a weighted average of the individual maps, where each pixel was attributed a Gaussian weight inversely proportional to the projected surface area it covers, and inversely proportional to its projected distance from the sub-Earth point.
Finally, we normalised the combined map with the global average albedo of Vesta in the Johnson V band (centred on 540~nm) in order to allow a direct comparison with the albedo map of \citet{Schroeder2014}.

The resulting map is shown and compared to that of \citet{Schroeder2014} in Fig.~\ref{fig:sph_map}. 
The Dawn map displays only the region of Vesta seen by SPHERE in order to facilitate the comparison. 
Other pixels were set to zero. 

The SPHERE map exhibits a wide range of albedo values (typically A$_{\rm V,N}$=0.34--0.45), close to that of Dawn (mostly A$_{\rm V,N}$=0.30--0.46), but slightly narrower owing to its lower spatial resolution and residual blurring due to imperfect deconvolution (Fig.~\ref{fig:histo_avn}). 
The peak of the two distributions is slightly offset, due to the different shape of the distributions: The SPHERE distribution appears even on both sides of the peak, whereas the low-end tail of the Dawn distribution is wider than the high-end one, possibly because multiple species with different albedos are spatially resolved.
Most of the main albedo features present in the Dawn map can also be identified in the SPHERE map. 
We find that surface features down to 20 km in size can be identified from the SPHERE map.

Only a few inconsistencies between the two maps are found. 
The most obvious example is the presence of a dark region located near $\lambda$=80$^{\circ}$, ${\phi}=-50^{\circ}$ on the SPHERE map, whereas Dawn finds no such albedo variation at this location.
This feature most likely results from the presence of shadows in the SPHERE images caused by irregular terrains rather than true albedo variations. 
The brightness of the southernmost region of Vesta on the map for $\lambda\in [180^\circ,270^\circ]$ and ${\phi}<-60^{\circ}$ is enhanced with SPHERE compared to Dawn. 
This is due to the illumination of the Rheasilvia central peak, which could not be accurately corrected. 
The Dawn map contains very localised regions with very low albedo values (A$_{\rm V,N}<$0.25). 
While these regions are also seen with SPHERE, they display higher albedos (A$_{\rm V,N}>$0.30) due to the lower spatial resolution of the map that attenuates the high-frequency albedo variations. 
Additional differences between the two maps may arise from the different filters used during the observations: ZIMPOL N\_R ($\lambda$=589.2--702.6~nm) versus DAWN FC2 clear filter ($\lambda$=438--965~nm).\\

\section{Conclusion}
\label{sec:conclusion}

In this article, we evaluated how much topographic and albedo information can be retrieved from the ground
with VLT/SPHERE in the case of the asteroid (4)~Vesta.
This object can be used as a benchmark for ground-based observations since the Dawn space mission provided us with ground truth information.
We observed (4)~Vesta with VLT/SPHERE/ZIMPOL as part of our ESO large programme at six
different epochs, and deconvolved the collected images with a parametric PSF. We then compared our
VLT/SPHERE images with synthetic images of Vesta produced with the OASIS software that uses the 3D shape model of the Dawn mission as input and on which we re-projected the albedo map of the Dawn mission \citep{Schroeder2014}. We further produced our own albedo map of Vesta from the SPHERE images alone, and without using any prior information about the surface topography of Vesta. We compared this map to that of \citet{Schroeder2014} to evaluate our ability to differentiate true albedo variegation from shadows and regions with enhanced illumination due to imperfect photometric correction of the SPHERE images. 
\\

We show that the deconvolution of the VLT/SPHERE images with a parametric PSF allows the retrieval of the main topographic and albedo features present across Vesta's surface down to a spatial resolution of $\sim$20--30 km. Contour extraction shows a precision of $\sim$1~pixel, with a strong difference between the estimation of the limb (precision of $\sim$0.5~pixel) and the terminator (precision of $\sim$1~pixel). The consequent relative error on the estimated area is $\leq 2\%$.\\

The present study demonstrates for the very first time the accuracy of ground-based AO imaging observations of asteroids with respect to {in situ} observations. Future generation telescopes (ELT, TMT, GMT) could use Vesta as a benchmark to catch all of the main features present across its surface (including the troughs and the north--south crater dichotomy), provided that these telescopes operate at the diffraction limit. \\



\textbf{Data retrieval}: As soon as papers for our large programme are accepted for publication, we will make the corresponding reduced and deconvolved AO images and 3D shape models publicly available at http://observations.lam.fr/astero/.

\begin{acknowledgements}
      This work was supported by the French Direction Générale de l'Armement (DGA) and Aix-Marseille Université (AMU). P.~Vernazza, A.~Drouard, and B.~Carry were supported by CNRS/INSU/PNP. J.~Hanu{\v s} was supported by the grant 18-09470S of the Czech Science Foundation and by the Charles University Research Programme No. UNCE/SCI/023. This project has received funding from the European Union’s Horizon 2020 research and innovation programme under grant agreement No. 730890. This material reflects only the authors’ views and the Commission is not liable for any use that may be made of the information contained herein. The authors thank S.~Schr{\"o}der for providing his reconstructed albedo map based on the Dawn images and for his very fruitful comments as a referee.

\end{acknowledgements}

\bibliographystyle{aa.bst} 
\bibliography{references.bib}



\begin{appendix}

\section{Observation table}

\begin{table*}
\caption{\label{tab:ao}List of Vesta images. For each observation, the table gives the epoch, the airmass, the distance to the Earth $\Delta$ and the Sun $r$, the phase angle $\alpha$, and the angular diameter $D_\mathrm{a}$. All observations were performed on the VLT/SPHERE instrument, with N\_R filter, and 80 seconds of exposure time.}
\centering
\begin{tabular}{rrr rrr r}
\hline
\multicolumn{1}{c} {Date} & \multicolumn{1}{c} {UT} &  \multicolumn{1}{c} {Airmass} & \multicolumn{1}{c} {$\Delta$} & \multicolumn{1}{c} {$r$} & \multicolumn{1}{c} {$\alpha$} & \multicolumn{1}{c} {$D_\mathrm{a}$} \\
\multicolumn{1}{c} {} & \multicolumn{1}{c} {} & \multicolumn{1}{c} {} & \multicolumn{1}{c} {(AU)} & \multicolumn{1}{c} {(AU)} & \multicolumn{1}{c} {(\degr)} & \multicolumn{1}{c} {(\arcsec)} \\
\hline\hline
  2018-05-20 &      6:41:53 & 1.01 & 1.25 & 2.15 & 15.8 & 0.579 \\
  2018-05-20 &      6:43:24 & 1.01 & 1.25 & 2.15 & 15.8 & 0.579 \\
  2018-05-20 &      6:44:56 & 1.01 & 1.25 & 2.15 & 15.8 & 0.579 \\
  2018-05-20 &      6:46:25 & 1.01 & 1.25 & 2.15 & 15.8 & 0.579 \\
  2018-05-20 &      6:47:56 & 1.01 & 1.25 & 2.15 & 15.8 & 0.579 \\
  2018-06-08 &      5:27:05 & 1.01 & 1.16 & 2.15 & 6.8 & 0.624 \\
  2018-06-08 &      5:28:36 & 1.01 & 1.16 & 2.15 & 6.8 & 0.624 \\
  2018-06-08 &      5:30:08 & 1.01 & 1.16 & 2.15 & 6.8 & 0.624 \\
  2018-06-08 &      5:31:38 & 1.00 & 1.16 & 2.15 & 6.8 & 0.624 \\
  2018-06-08 &      5:33:07 & 1.00 & 1.16 & 2.15 & 6.8 & 0.624 \\
  2018-06-08 &      7:14:40 & 1.08 & 1.16 & 2.15 & 6.7 & 0.624 \\
  2018-06-08 &      7:16:10 & 1.09 & 1.16 & 2.15 & 6.7 & 0.624 \\
  2018-06-08 &      7:17:42 & 1.09 & 1.16 & 2.15 & 6.7 & 0.624 \\
  2018-06-08 &      7:19:12 & 1.09 & 1.16 & 2.15 & 6.7 & 0.624 \\
  2018-06-08 &      7:20:41 & 1.10 & 1.16 & 2.15 & 6.7 & 0.624 \\
  2018-06-08 &      7:58:51 & 1.19 & 1.16 & 2.15 & 6.7 & 0.624 \\
  2018-06-08 &      8:00:22 & 1.19 & 1.16 & 2.15 & 6.7 & 0.624 \\
  2018-06-08 &      8:01:51 & 1.20 & 1.16 & 2.15 & 6.7 & 0.624 \\
  2018-06-08 &      8:03:21 & 1.20 & 1.16 & 2.15 & 6.7 & 0.624 \\
  2018-06-08 &      8:04:52 & 1.21 & 1.16 & 2.15 & 6.7 & 0.624\\
  2018-07-10 &      1:59:52 & 1.04 & 1.19 & 2.16 & 10.9 & 0.608 \\
  2018-07-10 &      2:01:24 & 1.03 & 1.19 & 2.16 & 10.9 & 0.608 \\
  2018-07-10 &      2:02:56 & 1.03 & 1.19 & 2.16 & 10.9 & 0.608 \\
  2018-07-10 &      2:04:25 & 1.03 & 1.19 & 2.16 & 10.9 & 0.608 \\
  2018-07-10 &      2:05:56 & 1.03 & 1.19 & 2.16 & 10.9 & 0.608 \\
  2018-07-10 &      2:16:38 & 1.02 & 1.19 & 2.16 & 10.9 & 0.608 \\
  2018-07-10 &      2:18:09 & 1.02 & 1.19 & 2.16 & 10.9 & 0.608 \\
  2018-07-10 &      2:19:39 & 1.02 & 1.19 & 2.16 & 10.9 & 0.608  \\
  2018-07-10 &      2:21:09 & 1.02 & 1.19 & 2.16 & 10.9 & 0.608  \\
  2018-07-10 &      2:22:39 & 1.02 & 1.19 & 2.16 & 10.9 & 0.608\\
\hline
\end{tabular}
\end{table*}

\section{Additional figure}
\label{sec:appendix_deconv}

\begin{figure*}
   \centering
   \includegraphics[width=16cm]{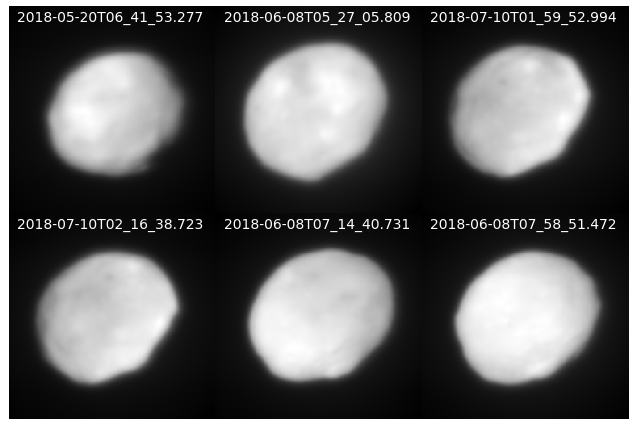}
   \includegraphics[width=16cm]{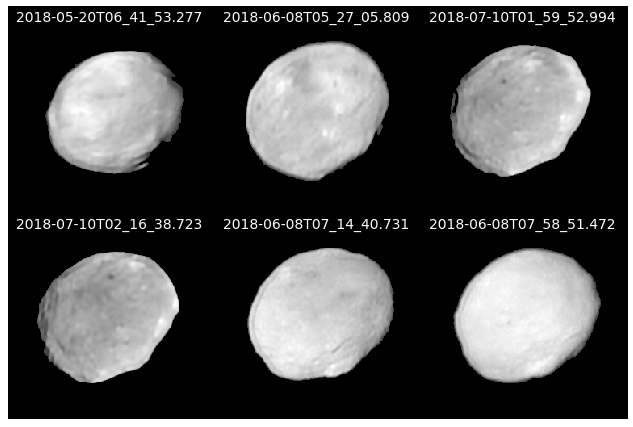}
   \caption{Top: Reduced Vesta images after pipeline. Bottom: Same images after deconvolution.}
\label{fig:VESTA_RAW}
\end{figure*}

\section{Optimisation of the PSF parameters}
\label{sec:appendix_optim}

\begin{figure*}
   \centering
   \includegraphics[width=18cm]{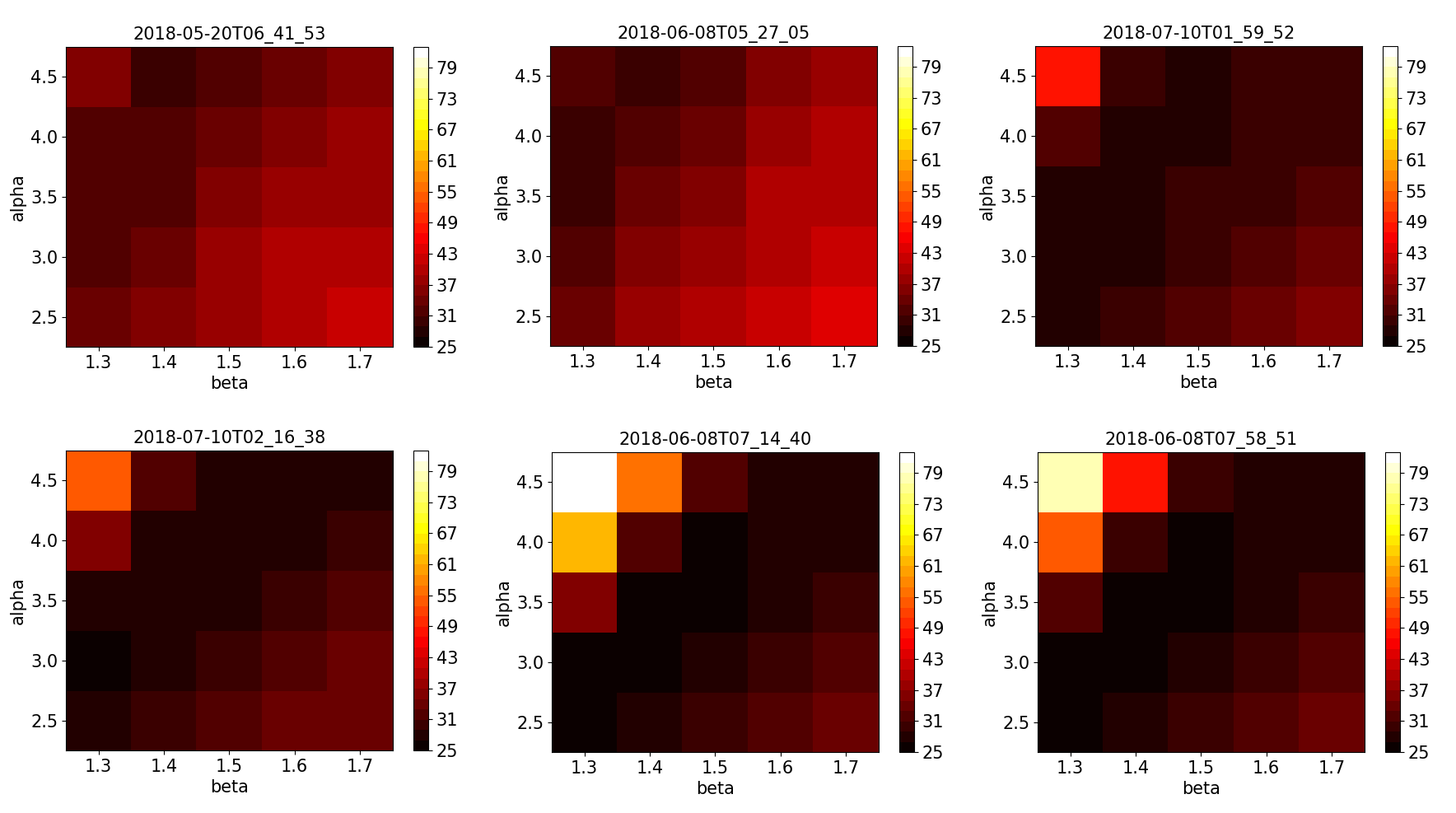}
   \caption{Root square error between deconvolved images and OASIS+albedo model. Each of the six subfigures corresponds to one of the observing epoch. Evolution of the error with respect to the $\alpha$ and $\beta$ parameters is then visible. For each epoch, optimal values of $\alpha$ and $\beta$ are the ones that minimise the error. Invalid pixels (missing albedo) were removed for error computation. However all pixels (inside and outside the asteroid) are considered to take into account the residual blurring after eventual incomplete deconvolution. Errors have also been divided by the flux of each Vesta observation so they reach comparable values between the different epochs.}
\label{fig:EQM}
\end{figure*}

Using OASIS synthetic images with albedo, we were also able to determine the optimal values of $\alpha$ and $\beta$ for the deconvolution by calculating the root square error between the deconvolved SPHERE/ZIMPOL images and the synthetic images (OASIS+albedo model). The error writes as
\begin{equation}
\label{eq:EQM}
    \epsilon(\alpha,\beta) = \sqrt{\sum_p (\text{DEC}_p(\alpha,\beta)-a_p\cdot\text{OAS}_p)^2}
,\end{equation}
where the sum runs over the pixels, OAS is the OASIS model multiplied by the albedo $a_p$ corresponding to each pixel, and DEC is our deconvolved image. When writing $\text{DEC}_p(\alpha,\beta)$ we recall that the deconvolution depends on the PSF parameters $(\alpha,\beta)$. The minimisation of the error $\epsilon(\alpha,\beta)$  identifies the optimal values for the $\alpha$ and $\beta$ parameters and thus the most accurate deconvolution in the sense of the norm given by Eq.\ref{eq:EQM}. Figure \ref{fig:EQM} shows the evolution of $\epsilon$ with respect to the PSF parameters $\alpha$ and $\beta$ for the six epochs. The minimum is reached in the narrow range $\alpha\simeq3.5 - 4$ and $\beta\simeq 1.5 - 1.6$. Figure \ref{fig:EQM} also shows that the error increases dramatically when the corona artefact appears, and that the minimum error is reached for sharp deconvolutions just before the corona effect.\\

Thus, the $(\alpha,\beta)$ parameters empirically estimated in Sect. \ref{subsec:parametricPSF} are close to the actual optimal values that minimise the $\epsilon$ criteria. Near such optimal values, only the corona effect may corrupt the deconvolution result at high $\alpha$ and low $\beta$. Otherwise the deconvolution zone is stable since the deconvolution quality evolves continuously with the parameters (see Figs. \ref{fig:deconv_range} and \ref{fig:EQM}).\\

Importantly, we believe that it is possible to further improve the deconvolution procedure using even more accurate PSF models, with a shape closer to that of the AO corrected PSF (paper in preparation on the AO PSF modelling). Finally, PSF parameters (whatever  PSF model is used) may be estimated automatically by a myopic deconvolution algorithm, such as  \citet{mugnier2004mistral} and \citet{Blanco-a-11}.

\begin{figure*}
   \centering
   \includegraphics[width=18cm]{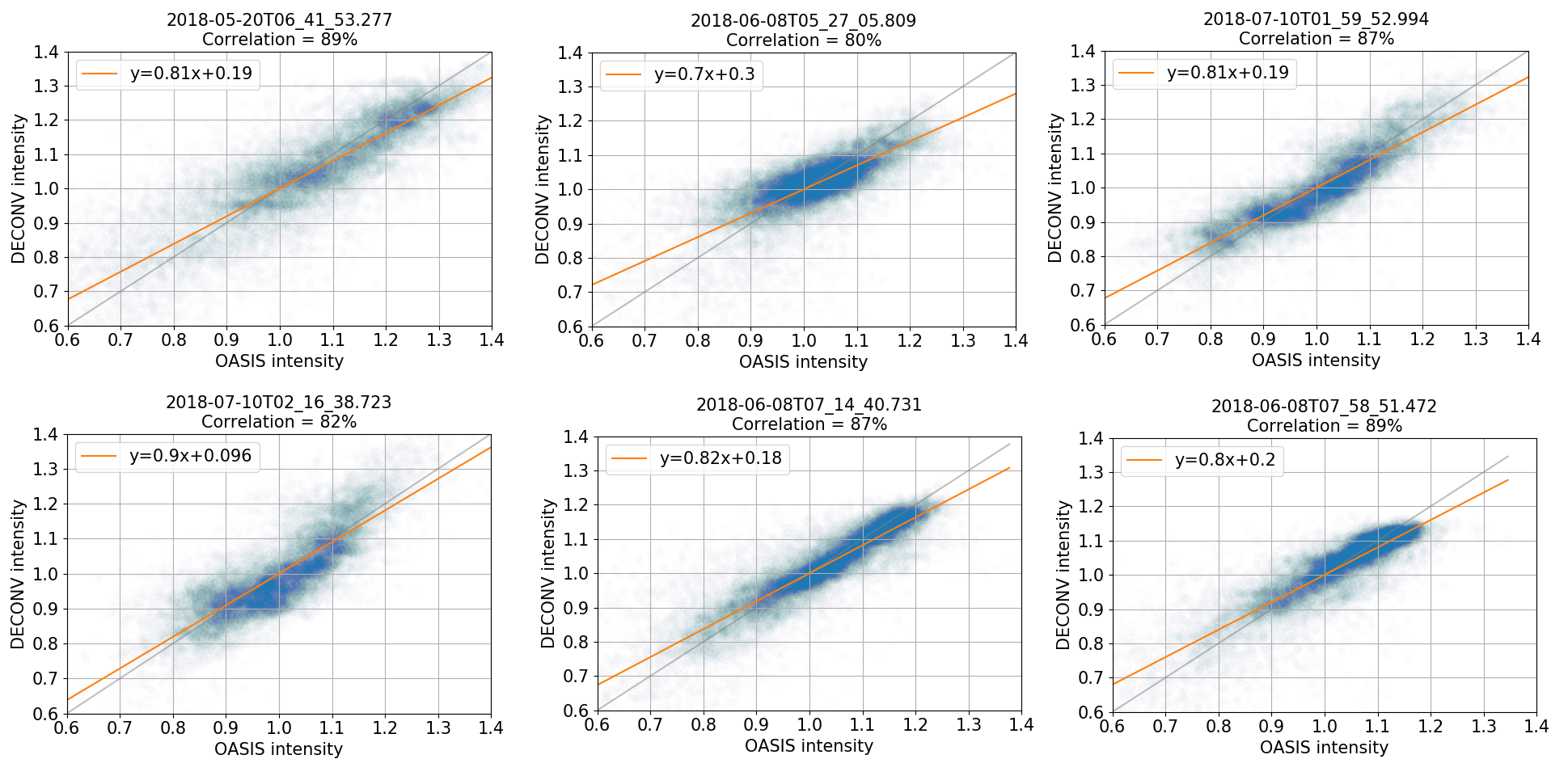}
   \caption{Correlation between OASIS (including albedo) and deconvolved image pixel intensities. Each intensity map has been divided by its average value for normalisation. Each of the six graphs corresponds to an epoch of observation. The linear fit of each cloud of points is also shown (orange line). The correlation percentage between OASIS and deconvolution intensities is written at the top of each graph.}
\label{fig:ALBEDO_COMP}
\end{figure*}

\end{appendix}


\end{document}